\documentclass[useAMS,usenatbib]{mn2e}

\def\msun{\hbox{M$_\odot$}}

\def\t4{\hbox{t$_{\rm 4}$}}

\def\three{\,{\sc iii}}

\usepackage{graphicx}
\usepackage{natbib}
\usepackage{amssymb}
\usepackage{lscape}
\usepackage{longtable}
\title[A lack of ongoing star formation within YMCs]{Constraining Globular Cluster Formation Through Studies of Young Massive Clusters:  I.  A lack of ongoing star formation within young clusters}
\author[N. Bastian et al.]{ N. Bastian$^{1}$\thanks{NB: N.J.Bastian@ljmu.ac.uk}, I. Cabrera-Ziri$^{1,2}$, B. Davies$^{1}$, S.S. Larsen$^3$\\
$^{1}$ Astrophysics Research Institute, Liverpool John Moores University, 146 Brownlow Hill, Liverpool L3 5RF, UK\\
$^{2}$ Centro de Investigaciones de Astronom\'ia, Av. Alberto Carnevalli, A.P. 264, C.P. 5101, M\'erida, Venezuela\\
$^{3}$ Department of Astrophysics / IMAPP, Radboud University Nijmegen, PO Box 9010, 6500 GL Nijmegen, the Netherlands\\
}
\begin{document}

\date{Accepted XXX. Received XXX; in original form XXX}

\pagerange{\pageref{firstpage}--\pageref{lastpage}} \pubyear{2013}

\maketitle

\label{firstpage}

\begin{abstract}
We present a survey of 130 Galactic and extragalactic young massive clusters (YMCs, $10^4 < M/\msun < 10^8$, $10 < t/{\rm Myr} < 1000$) with integrated spectroscopy or resolved stellar photometry (40 presented here and 90 from the literature) and use the sample to search for evidence of ongoing star-formation within the clusters.  Such episodes of secondary (or continuous) star-formation are predicted by models that attempt to explain the observed chemical and photometric anomalies observed in globular clusters as being due to the formation of a second stellar population within an existing first population.  Additionally, studies that have claimed extended star-formation histories within LMC/SMC intermediate age clusters (1-2~Gyr), also imply that many young massive clusters should show ongoing star-formation.   Based on visual inspection of the spectra and/or the colour-magnitude diagrams, we do not find evidence for ongoing star-formation within any of the clusters, and use this to place constraints on the above models.  Models of continuous star-formation within clusters, lasting for hundreds of Myr, are ruled out at high significance (unless stellar IMF variations are invoked).  Models for the (nearly instantaneous) formation of a secondary population within an existing first generation are not favoured, but are not formally discounted due to the finite sampling of age/mass-space. 

\end{abstract}


\section{Introduction}
\label{sec:intro}

Once thought to be the quintessential simple stellar populations (i.e. all stars having the same age and chemical composition), the ancient globular clusters (GCs) in the halo/bulge of our Galaxy are now known to host a number of anomalies. In particular, the stars within GCs display a number of (anti)correlations in the abundance patterns of light elements (e.g., the Na-O anti-correlation), showing that abundance variations are present within clusters (e.g., Gratton et al. 2012 and references therein). Interestingly, these trends are only found in a very small fraction ($\sim3$\%) of old field stars in the halo (Martell et al. 2011), and these stars may have been lost from GCs. This implies that it is only within the unique environment of GCs that such anomalies can develop. Additionally, high precision colour-magnitude diagrams (CMDs) of GCs have found that nearly all GCs display complex and unexpected features, such as double main sequences, extended horizontal branches and multiple giant/sub-giant branches (e.g., Piotto et al. 2012). 

A number of theories have been put forward to explain the observed anomalies,  most of which invoke the formation of a second stellar population within an existing massive cluster (the first generation) from material processed in the first stellar generation  (e.g., Decressin et al. 2007; D'Ercole et al. 2008; de Mink et al. 2009).   These theories, while explaining many of the chemical and photometric anomalies, all have a ``mass budget problem", and require the first generation to have been significantly more massive than observed today, in conflict with observations of the nearby Fornax dwarf galaxy (Larsen, Strader, \& Brodie~2012).

In addition, observations of intermediate age clusters ($1-2$~Gyr) in the Large and Small Magellanic Clouds have shown the presence of extended main-sequence turn-offs (eMSTOs) that are not consistent with a single theoretical isochrone (e.g., Mackey \& Broby-Neilsen~2007).  This has been interpreted as evidence for a significant age spread present within the clusters, with a duration of $200-500$~Myr (Milone et al.~2009, Goudfrooij et al.~2009, 2011a,b; Rubele et al. 2013).  Additionally, some clusters have 'dual red clumps' in their CMDs which can be interpreted as an age spread (e.g., Girardi et al. 2009). A number of works have attempted to link these implied age spreads with the anomalies in the ancient GCs (e.g., Conroy \& Spergel~2011, Keller, Mackey, \& Da Costa~2011), suggesting a common evolution for massive clusters (present day masses $\gtrsim 10^5$\msun). One potential caveat with these models is that the inferred age spreads in LMC/SMC cluster are continuous (e.g., Goudrooij et al. 2011b), while the CMD morphology of many GCs suggest discrete events.  Based on empirical relations, Goudfrooij et al. (2011b) suggest that only clusters with escape velocities greater than $10-15$~km/s should be able to have multiple/extended episodes of star-formation, as they suggest that only these can retain the ejecta of evolving stars.

Models that interpret the chemical abundance anomalies in GCs and/or the complex CMDs of GCs and intermediate age clusters as multiple star formation episodes, all make a common prediction, that young massive clusters (YMCS; $<500$~Myr) should show ongoing star-formation as well as gas/dust build-up within them that will eventually be used in the formation of a second generation.  The duration of the age spread (or time between successive discrete star forming episodes) depends on the model.  If the second generation of GCs forms from the enriched material ejected from rapidly rotating massive stars (e.g., Decressin et al. 2007) or interacting binaries (e.g., de Mink et al. 2009), then the expected time between the first and second generation is expected to be short ($\sim5-20$~Myr).  On the other hand, if AGB stars are the source of the second generation, then the time difference is expected to be $30-200$~Myr (e.g., D'Ercole et al. 2008; Conroy \& Spergel~2011).  If the eMSTOs are due to extended SFHs, then young clusters should be seen forming stars continuously for $200-500$~Myr.

Observations of young massive ($1-10 \times 10^4$\msun) clusters in the Galaxy, such as Westerlund~1 and NGC~3603, suggest an upper limit to age spreads of $1-2$~Myr (e.g., Kudryavtseva et al.~2012).  Additionally, the clusters appear to be devoid of dense gas at an early age, $<5$~Myr (e.g., Muno et al. 2006).  These timescales are too short for any of the mechanisms proposed for the multiple populations of GCs to operate.  However, these clusters may be too low mass for the mechanisms to work.  Conroy \& Spergel~(2011) predict that clusters, at least in the LMC, with masses in excess of $10^4$\msun\ should be able to retain processed/enriched material and accrete new material from the surroundings in order to form a second generation of stars.  Bastian \& Silva-Villa~(2013), however, found no such extended star-formation histories in two massive ($10^5$~\msun), relatively young (180 and 280~Myr) clusters in the LMC, in conflict with the predictions.

We note that alternatives explanations to both phenomena have been put forward.  Stellar rotation and/or interacting binaries may be able to explain the eMSTOs seen in LMC/SMC clusters (Bastian \& de Mink 2009; Yang et al.~2011 - although see Girardi, Eggenberger, \& Miglio~(2011) and Yang et al.~(2013) for a further discussion on the effects of stellar rotation), although this explanation will not work for the ancient GCs.  Bastian et al. (2013) have recently put forward a theory for the observed anomalies in GCs that does not require multiple episodes of star-formation within clusters.  Instead, it invokes the enrichment of a fraction of low-mass stars through disc accretion within the cluster from the ejecta of high mass stars, a process that is predicted to be operating in all clusters above a critical density.

There are clusters in the local universe that have masses similar to, and even significantly above, GCs, that can be used to directly test the above scenarios.  These young massive clusters (see, e.g., Portegies Zwart, McMillan \& Gieles~2010 for a recent review) have ages between a few Myr and a few hundred Myr, and masses between $10^4$ and $\sim10^8$\msun\ (e.g., Maraston et al. 2004), although we note that the lower limit is a relatively arbitrary cut as the mass function of clusters extends smoothly below this limit.  Over the past two decades, a significant database of the properties of these clusters has been developed.  In the present work, we will concentrate on those clusters that have integrated spectroscopy  or resolved stellar photometry.  Following on the work of Peacock et al.~(2013), we ask the relatively simple and straightforward question, do any of these clusters show evidence for ongoing star formation within them?  In a future work, we (Cabrera-Ziri et al.~in prep.) will look in more detail at what constraints can be put on the SFH of clusters from their integrated spectra, which will more directly constrain the presence of multiple star formation events within clusters.


This paper is organised as follows: in \S~\ref{sec:obs} we present new spectroscopy for a sample of clusters in four spiral galaxies.  In \S~\ref{sec:catalogue} we supplement this sample with a collection of clusters taken from the literature, and use the combined catalogue to search for evidence of ongoing star-formation within the clusters, and explore in detail a few interesting (high mass) cases.  In \S~\ref{sec:constraints} and \S~\ref{sec:models} we will explore the resulting constraints on the proposed models that are obtained with the observations and in \S~\ref{sec:discussion} we discuss the implications of our results and present our conclusions.


\section{EMMI spectra}
\label{sec:obs}

We include spectra of a number of young clusters taken with the EMMI
(ESO Multi-Mode Instrument) spectrograph at the 3.5 m New
Technology Telescope at ESO, La Silla. These observations were made
on March 16 - 18, 2004. The clusters were mostly selected from
the list in Larsen (2004), with emphasis on clusters that appeared
well-defined, regular and isolated in HST images.  The spectra cover 
the wavelength range
4000--9000 \AA\ at a resolution of $\lambda/\Delta \lambda \sim 760$
and were obtained with the RILD ("Red Imaging and Low-Dispersion
Spectroscopy") mode of EMMI in multi-slit mode. A total of four
galaxies were observed, using three slitmasks for NGC~5236, two
for NGC~3621 and one each for NGC~2835 and NGC~2997. Exposure times
were typically $3\times60$ min per slitmask.
The spectra were reduced with standard tasks in IRAF. Due to the
lack of an atmospheric dispersion corrector and the constraints on
the orientation of the slits imposed by the slitmask design,
wavelength-dependent slitlosses were significant. An approximate
flux calibration was done by scaling the spectra to BVRI
broad-band photometry of the clusters (Larsen 1999).
For more details we refer to Larsen \& Richtler (2006).

\begin{figure*}
\includegraphics[width=15cm]{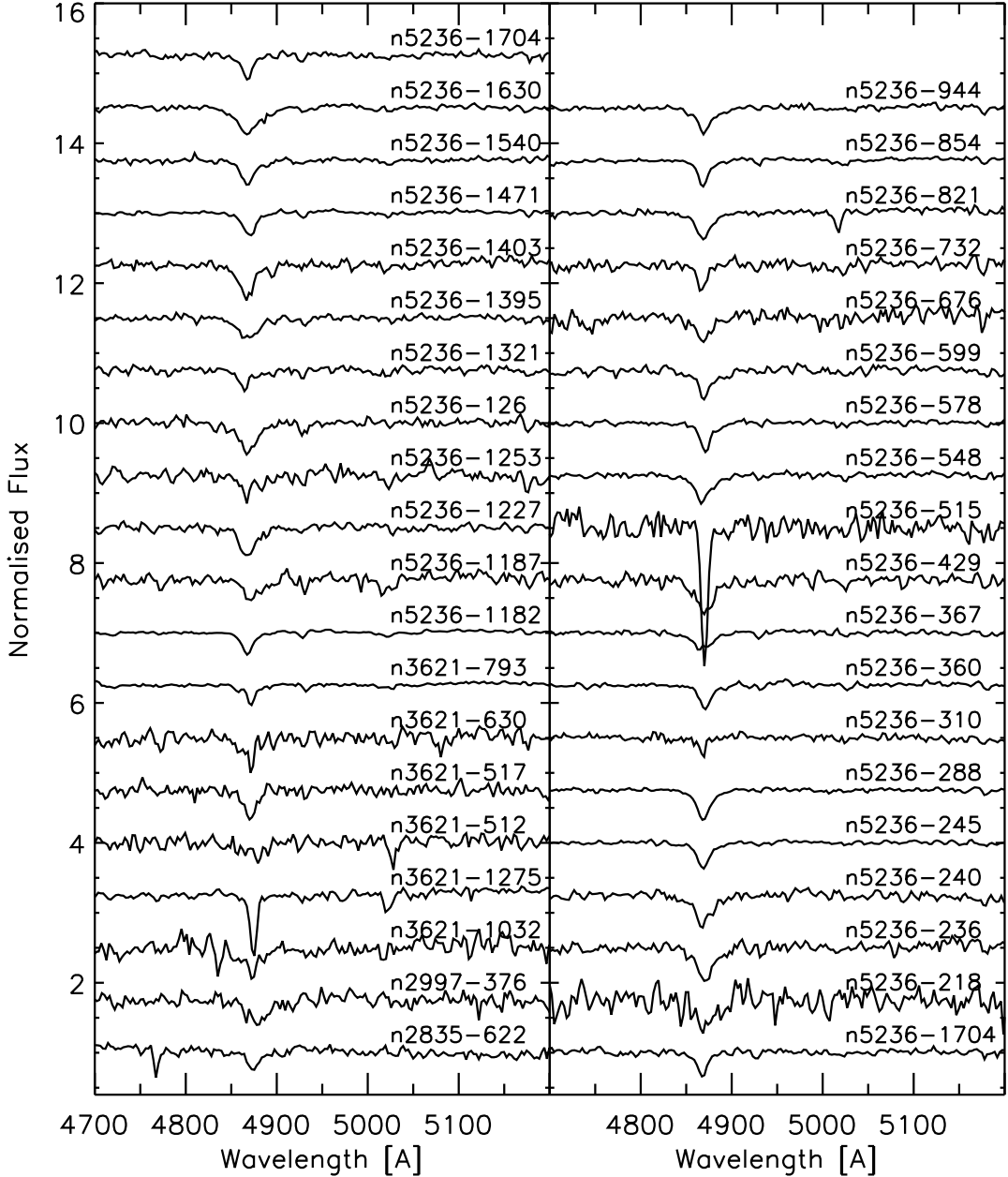}
\caption{Integrated spectra of the clusters presented in this paper (\S~\ref{sec:obs}).  We focus on the region of the spectrum that contains three prominent emission lines, H$\beta$, O[\three]4959 and O[\three]5007, in order to look for evidence of ongoing star formation within the clusters. The cluster IDs are given above each spectrum.  All spectra are shown in observed wavelength.}
\label{fig:larsen_spectra}
\end{figure*}

Ages and masses of the clusters were estimated from the UBVRI broad-band
photometry via comparison with the simple stellar population models of
Bruzual \& Charlot (2003). We used Solar-metallicity models, computed
for a Chabrier (2003) IMF with a lower mass limit of 0.1 M$_\odot$.  The best fitting model was found by including the effects of interstellar extinction in the models and minimising the reduced $\chi^2$.
A total of 80 clusters were observed. After restricting the age range
to ages between 10 Myr and 1 Gyr, and removing clusters with low S/N (those where H$\beta$ was not detected in absorption), 40 objects remained.  The normalised spectra of the 39 clusters used in the present work are shown in Fig.~\ref{fig:larsen_spectra}, centred on the H$\beta$ and O[{\sc iii}]$\lambda$5007 lines, to search for emission associated with the clusters.

One cluster in this sample, NGC~5236-693 (age $\approx 36$~Myr , mass $\approx 9 \times 10^4$) from the Larsen \& Richtler~(2006) deserves special note (and is not shown in Figure~\ref{fig:larsen_spectra}). While the integrated spectrum of this  cluster has  H$\beta$ in emission superposed on an absorption component, analysis of available HST images shows that the emission is due to a nearby H{\sc ii} region, and does not originate from the cluster itself.  Two other clusters from the literature (T111 in the Antennae and T661 in NGC~3256) are similar and are discussed in more detail in \S~\ref{sec:notes} (see also Peacock et al. 2013).

\section{YMCs taken from the Literature}
\label{sec:catalogue}

The catalogue used in the current work has been collected from a number of studies in the literature.  We searched for clusters with masses above $10^4$\msun, and ages greater than 10~Myr that have published integrated spectroscopy or resolved stellar photometry.  The mass limit was adopted to avoid the complications of a stochastically sampled stellar initial mass function (e.g., Fouesneau 
\& Lan{\c c}on~2010 and Silva-Villa \& Larsen 2010).  Additionally, most studies that have claimed the presence of age spreads (or have modelled when they should exist) suggest that they should only be present in massive clusters (e.g., Goudfrooij et al. 2011b).  The lower age limit was applied in order to avoid clusters that have gaseous emission associated with them, which is not linked to ongoing star-formation (i.e. the clusters have high mass stars, from the initial formation episode, within them that are ionising surrounding gas).  Hence, we can not test age spreads which are shorter than this lower age limit, meaning that we cannot test the predictions of the rapidly rotating or interacting binary models for globular clusters, which act on similar or shorter timescales.   

\begin{figure}
\includegraphics[width=9cm]{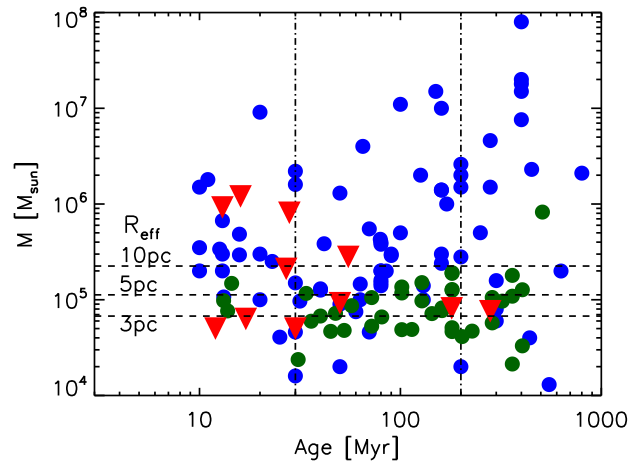}
\caption{The age-mass diagram of the clusters in our sample.  Filled blue and green circles represent clusters studied through integrated spectroscopy taken from the literature and from the current study, respectively.  Upside down (red) filled triangles represent clusters studied through resolved photometry (taken from the literature).  The vertical dash-dotted lines mark the upper/lower limits of when the "second generation" is expected to form in the AGB model for GCs (see \S~3.2.1).  The horizontal dashed lines represent the mass limit associated with an escape velocity of $15~km/s$, for effective radii of 3, 5 and 10~pc, from bottom to top, respectively (see \S~3.2.1).}
\label{fig:age_mass}
\end{figure}

The age, mass, ID, host galaxy and references of each cluster is given in Table~\ref{tab:clusters}.  We note that the ages of the clusters were estimated under the assumption that the clusters are simple stellar population (i.e. all stars have the same age within some small tolerance).  In a future work (Cabrera-Ziri et al., in prep), we will investigate the effect of this assumption with detailed spectral synthesis modelling of multiple populations. The age and mass of the clusters in our sample are shown in Fig.~\ref{fig:age_mass}, where filled blue and green circles represent clusters with integrated spectroscopy taken from the literature and presented here, respectively, and circles and (red) upside down triangles represent clusters studied with resolved photometry (also taken from the literature.  Typical errors on the estimated age and mass are a factor of two.

The majority of the spectra have a similar resolution (R = $1000-2000$) and we restrict our analysis to those cluster with high enough S/N ($\sim10$) in order to clearly detect the Balmer absorption lines.

A handful of clusters in our sample have been studied with high-resolution imaging with the Hubble Space Telescope, allowing the construction of colour-magnitude diagrams (CMDs) of the resolved stellar populations within the clusters.  These clusters are shown as red triangles in Fig.~\ref{fig:age_mass}. Through the analysis of the CMDs, the presence/absence of short-lived very massive main sequence stars (indicative of ongoing star-formation) is readily apparent, due to their brightness.  No such massive stars are seen in the clusters studied to date.  We note that some of the resolved clusters have complex CMDs (NGC~1569B, NGC~1705-1, NGC1313-F1-1, NGC~5236-F1-1, -F1-3, NGC~7793-F1-1) that may indicate extended star-formation histories, lasting for up to 30~Myr, although none show evidence for ongoing star formation, and interacting binaries may explain the CMD morphologies (Larsen et al.~2011).

\subsection{Note on individual clusters}
\label{sec:notes}

\subsubsection{M82 F and L}
\label{sec:m82f}

M82~F and L are two massive clusters in the nearby, edge-on starburst galaxy, M82.  Both are in (at least in projection) a region of high absolute and differential extinction.  Both clusters have been studied in detail, in multiple ground based integrated spectroscopic studies (Gallagher \& Smith~1999; Smith \& Gallagher~2001; McCrady \& Graham~2007; Konstantopoulos et al. 2009), although due to the lower extinction, M82F has been subjected to heavier scrutiny.  The integrated spectra of both clusters contain clear emission features (H$\beta$, O[{\sc iii}], H$\alpha$).  However, high resolution echelle spectroscopy (Gallagher \& Smith 1999; Smith \& Gallagher 2001) has shown that the emission components are offset in velocity, relative to both clusters, hence are not associated with either cluster.

\subsubsection{T111 in the Antennae}
The YMC T111 in the Antennae galaxies (Bastian et al. 2009) has recently been studied in detail by Peacock et al. (2013).  This cluster has all the expected attributes of a relatively old ($\sim80$~Myr) cluster with ongoing star-formation.  The cluster shows clear absorption lines in the integrated spectra, as well as strong emission lines, near the absorption line centres (in the case of the Balmer lines), indicating the presence of ionised gas (and presumably ionising stars) at a similar velocity as the older cluster.  However, upon closer inspection, the authors found that the emission lines are coming from a nearby young cluster ($<7$~Myr) that is not associated with T111.  Due to the small velocity difference between the two clusters the authors suggest that the clusters may merge.  This is definitely possible, and while an interesting route to the formation of multiple populations within a single (merged) cluster, it is beyond the scope of the current paper.

\subsubsection{T661 in NGC 3256}
\label{sec:t661}

\begin{figure}
\begin{center}
\includegraphics[width=5cm]{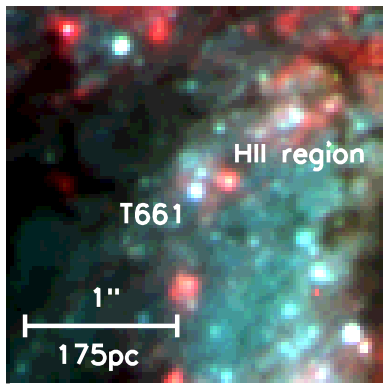}
\end{center}
\caption{A colour composite image of the cluster T661 in NGC~3256.  Blue, green and red represent the HST/ACS images in the F435W, F555W and F656N (H$\alpha$) filters respectively.  Note the strong H{\sc ii} region near T661, the likely origin of the line emission seen in the ground based spectrum.}
\label{fig:t661}
\end{figure}

T661 in the ongoing galactic merger NGC~3256 shows a similar spectrum as that observed in T111, i.e., an absorption component (suggestive of an age of $\sim50$~Myr) and an emission component superposed.  In Fig.~\ref{fig:t661} we show a colour composite image of T661 made with HST/ACS WFC3 F435W (B), F555W (V) and  F656N (H$\alpha$) imaging.  Like that found for T111 in the Antennae by Peacock et al. (2013), there is an H{\sc ii} region nearby T661 that would be included in the ground based spectroscopic slit.  Hence, we conclude that the emission seen in the integrated spectra is not due to star formation within T661, but is from the (unrelated) nearby H{\sc ii} region.  Like T111 in the Antennae, T661 may potentially be a binary cluster.


\subsection{Clusters with masses exceeding $10^6$\msun}

Of particular interest in the current work are the clusters with estimated masses exceeding $10^6$\msun.  The reason for this is that they are approximately a factor of ten more massive than a typical GC in the Galaxy today.  As multiple theories for the origin of the multiple populations of GCs require that they were ten times (or more) massive at birth than they are today, these YMCs are the best place to test the GC formation theories.

\begin{figure}
\includegraphics[width=8cm]{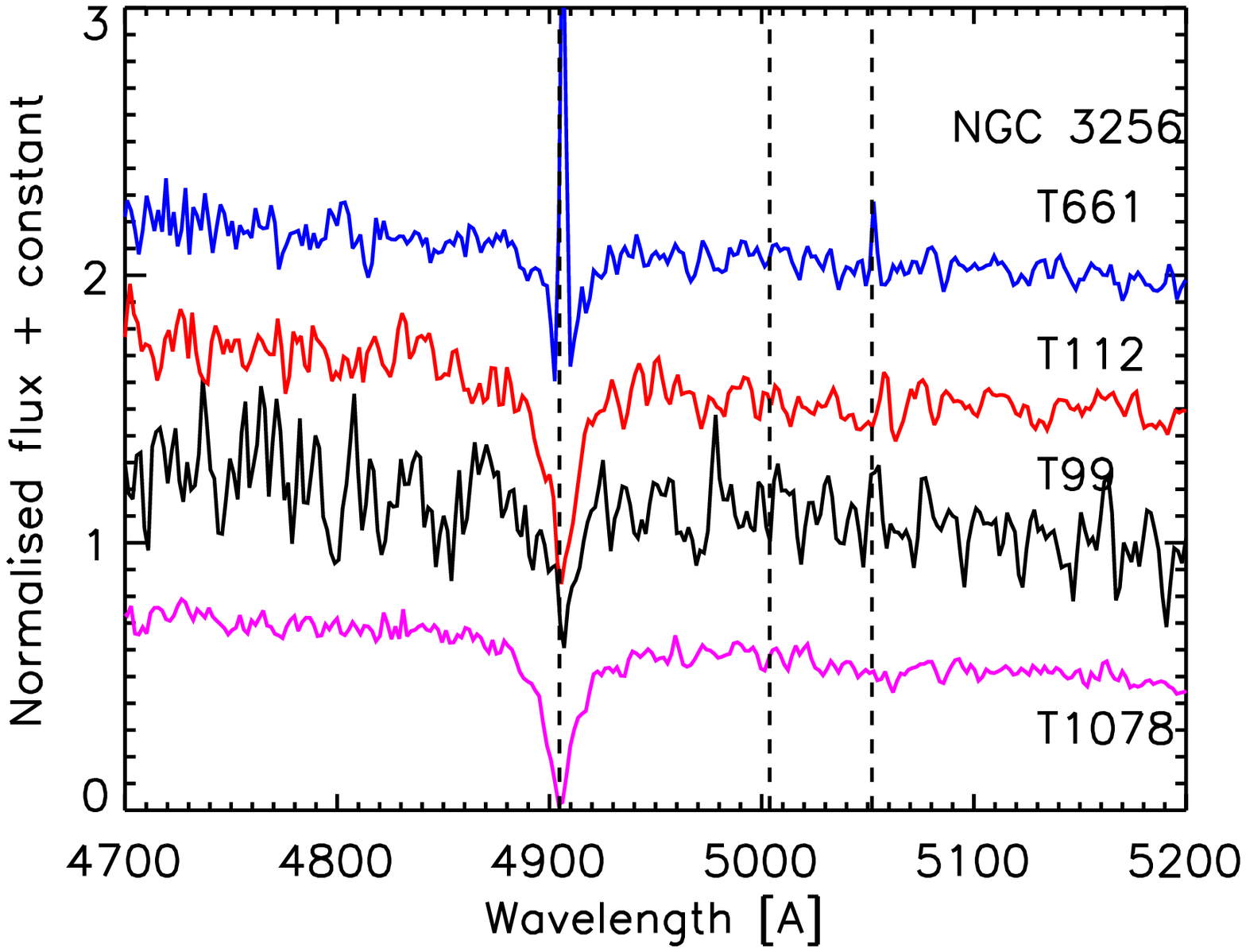}
\includegraphics[width=8cm]{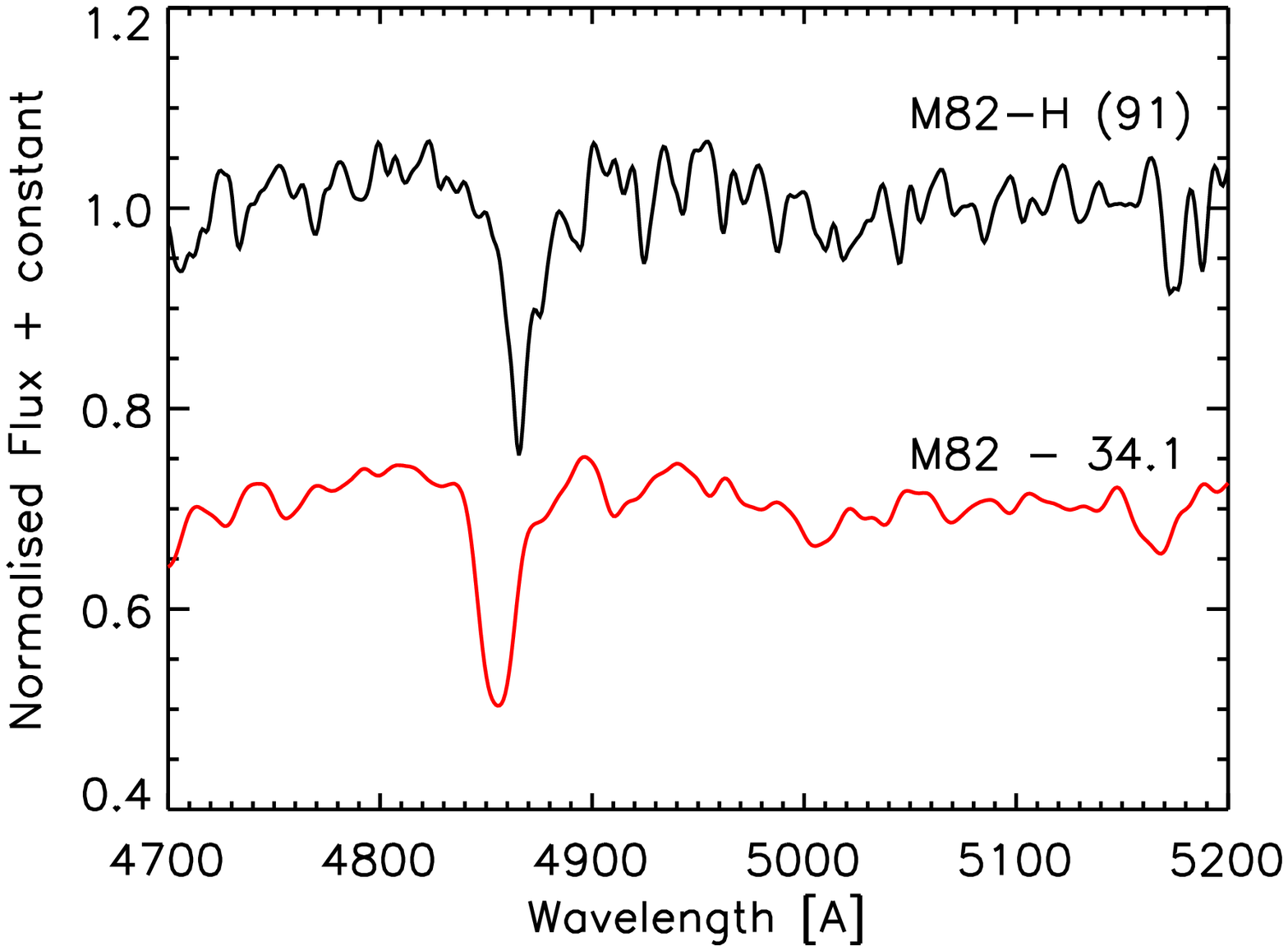}

\caption{{\bf Top:} Integrated optical spectra from Trancho et al. (2007b) of four young clusters in NGC~3256 with estimated masses in excess of $10^6$\msun.  The spectra are in the observed frame, and the vertical dashed lines indicate the expected position of H$\beta$, O[\three]4959 and O[\three]5007.  {\bf Bottom:} The same as the top panel, but now for two massive clusters in M82.  All spectra are displayed in observed wavelength.}
\label{fig:n3256_spec}
\end{figure}

In the top panel of Fig.~\ref{fig:n3256_spec} we show the integrated spectrum of four young clusters in NGC~3256 with estimated masses in excess of $10^6$\msun.  Only cluster T661 shows evidence for line emission.  However, as discussed above this is likely associated with a nearby H{\sc ii} region, and not associated with the cluster.  In the bottom panel of Fig.~\ref{fig:n3256_spec} we show the integrated spectra of two massive clusters in M82.  Neither shows evidence of associated nebular emission.

\subsection{Clusters not included}

Larsen \& Richtler (2006) found three objects that had emission associated with an underlying absorption component among the 80 YMCs with EMMI spectra (see \S~\ref{sec:obs}).  Due to the line ratios and strengths the authors conclude that the emission is not due to an underlying H{\sc ii} region, but rather is due to the presence of a planetary nebula (PNe) within each cluster.  The cluster IDs are NGC~5236-254, NGC~5236-487 and NGC3621-1106.  These clusters have estimated ages of 180, 32 and 50~Myr and masses of $2.7 \times10^5$, $2.5 \times 10^4$, and $8.7 \times 10^4$\msun, respectively.  

It is interesting to note that none of the additional clusters in the sample used here show PN features. Larsen \& Richtler~(2006) estimated that $\sim6$~PNe should have been found in their survey.  The authors note that the three PNe candidates were all found in relatively extended (R$_{\rm eff} > 6$~pc) clusters, perhaps indicating that the average PNe lifetime (normally assumed to be $\sim10^4$~yr) is significantly shorter in dense environments.  An alternative explanation is that the emission is due to symbiotic stars (i.e. a red giant transferring mass onto a white dwarf).

Due to the interpretation of the emission lines most likely stemming from PNe, we have not included these four objects in our sample. The other clusters in the Larsen \& Richtler~(2006) sample (with ages between 10 and 1000 Myr, and with high enough S/N to detect H$\beta$ in absorption) have been included and are presented in \S~\ref{sec:obs}.




\section{Constraints on ongoing star-formation}
\label{sec:constraints}

The presence/absence of emission lines in the integrated spectra of the clusters was determined through visual inspection.  In this section we attempt to quantify the constraints that the observations can provide, both in terms of the mass limit of any secondary star formation episode as well as the total amount of ionised gas that may be present.

In order to quantify what constraints can be put on the lack of emission coming from clusters, Peacock et al.~(2013) simulated the expected spectrum of a 100~Myr cluster with a 4~Myr population superimposed.  Both spectra were taken from the {\rm STARBURST99} simple stellar population library (Leitherer et al.~1999), which includes the expected contribution from nebular emission lines.  Based on their Figure~4 (left panel), if an older population (100 Myr) had any ongoing star formation this would be easily seen in the integrated spectrum, if the young population contributed more than 3\% of the total stellar mass of the system.  Extrapolating their results, we see that even if the secondary population was a few tenths of a percent of the mass of the older population, the emission lines would be readily apparent.  This is a best case scenario, as it assumes extremely high signal-to-noise.

We can test these results empirically, by summing the spectra of two clusters, one with prominent emission lines and one dominated by absorption lines, characteristic of an older population (a few tens of Myr).  For this experiment, we chose three clusters from the Antennae galaxies, studied by Bastian et al. (2009).  For the absorption line cluster, we use T296, an $\sim80$~Myr cluster with an estimated mass of $\sim4 \times 10^5$~\msun.  We experimented with two emission line clusters, T352 ($\sim5 \times 10^5$\msun) and T395 ($\sim2 \times 10^5$\msun), due to their different emission line ratios.  We refer the reader to Bastian et al. (2009) for a description on how the cluster properties were derived.  We corrected for their small mass differences but did not correct for extinction, given the uncertainty in the estimate, although we note that the extinction in all three is estimated to be low ($A_V \lesssim 0.3$).  We then summed the absorption and emission line cluster spectra together, dividing the flux of the emission line cluster by a given factor, ranging from 2 to 100.  This simulates the expected change if the young population was less massive than the older cluster.

\begin{figure}
\includegraphics[width=8.5cm]{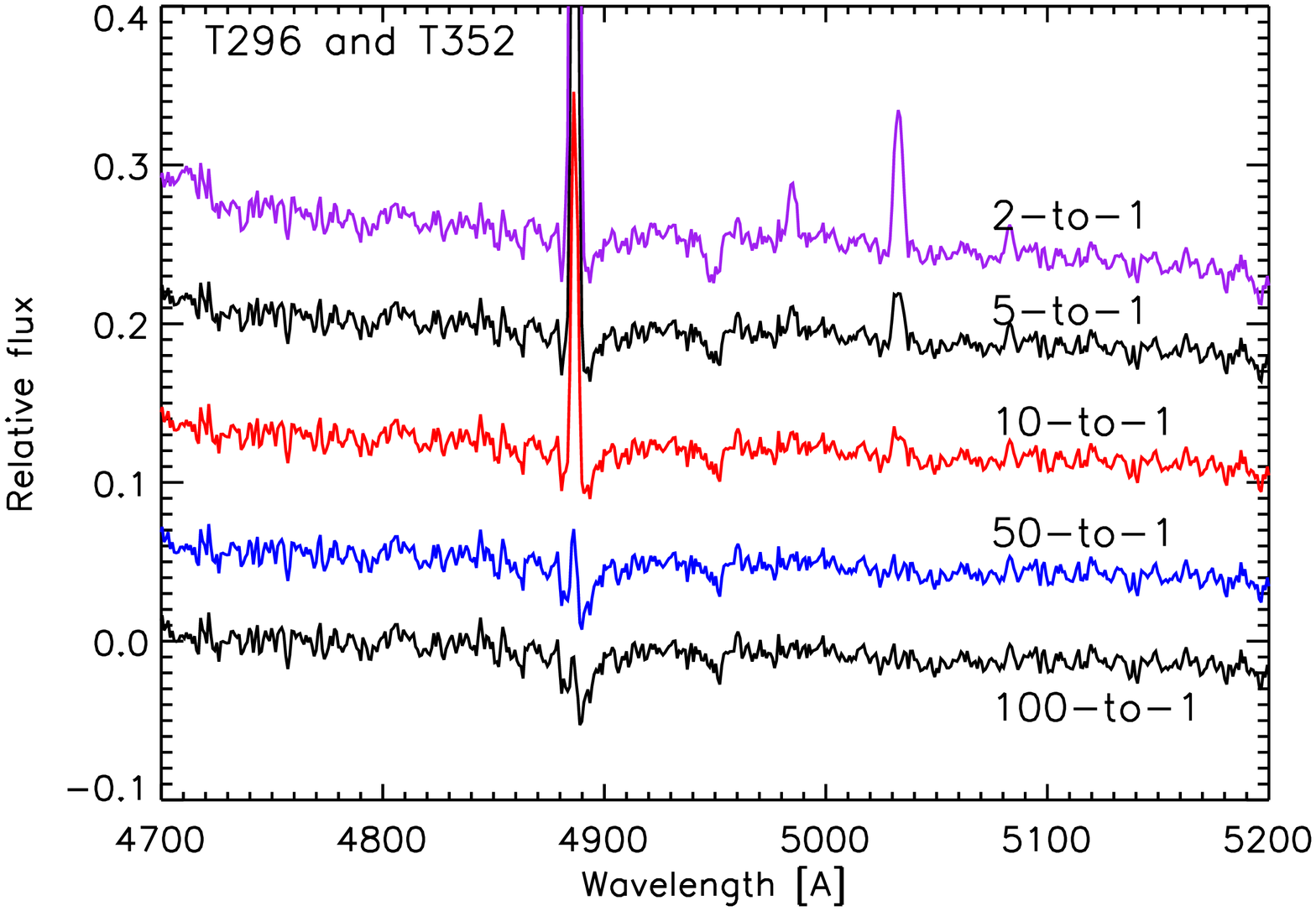}
\includegraphics[width=8.5cm]{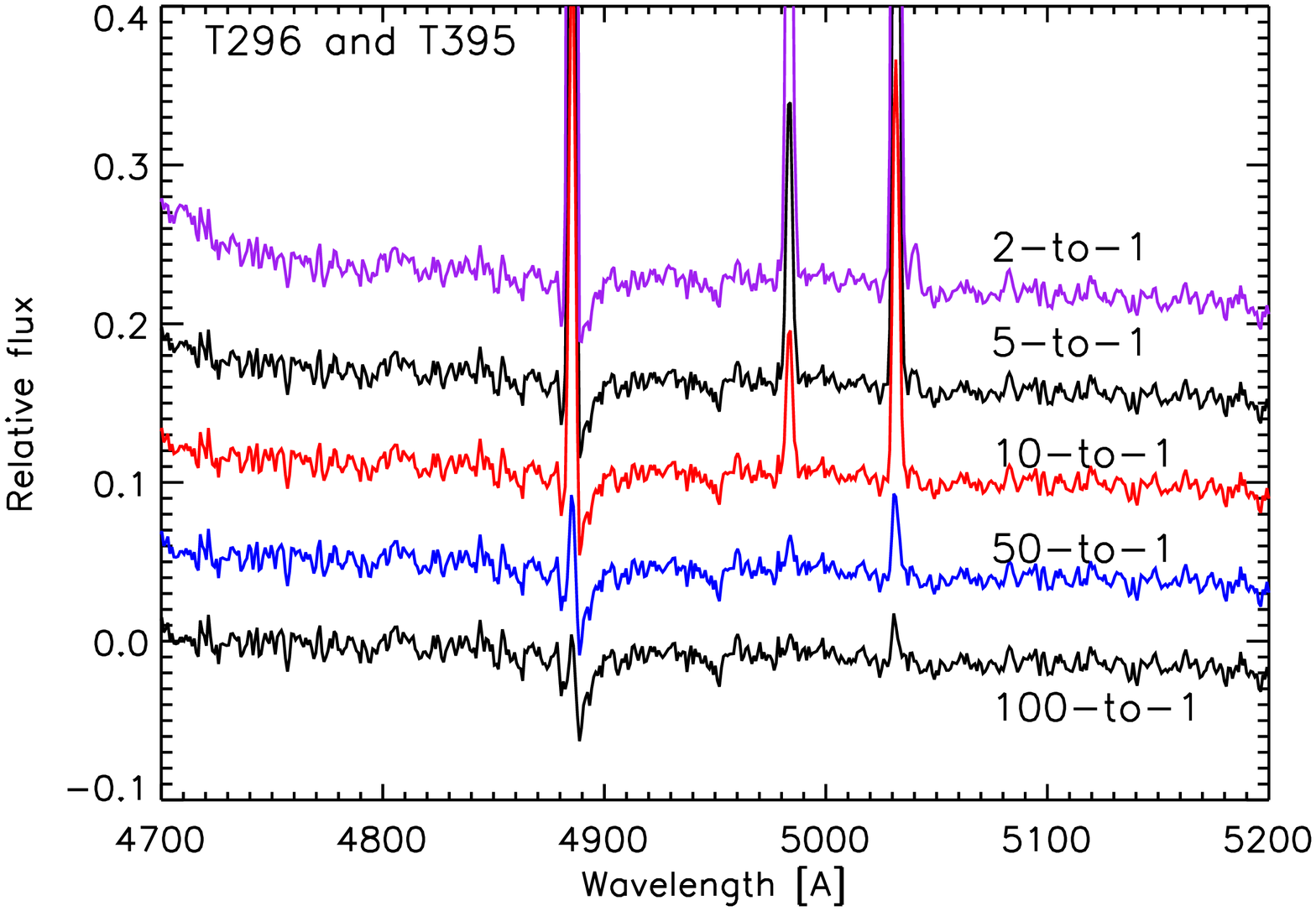}
\caption{An empirical determination of the limit to which a newly forming stellar population could be seen within a more massive, older cluster.  The two panels show the resulting spectra of an absorption and emission line cluster, taken from Bastian et al.~(2009), summed together at different mass ratios.  The mass ratio for each combination is given (with the young population always being less massive).  In both cases, a mass ratio of 50-to-1 would have been easily detected in H$\beta$ and/or O[\three].}
\label{fig:empirical}
\end{figure}

The results are shown in Fig.~\ref{fig:empirical}. We note that even at 2-to-1 (old-to-young) mass ratios, the underlying older population can be readily seen in H$\beta$ absorption.  Hence, while the presence of an older population may be difficult to discern photometrically, integrated spectra can place strong constraints on such a presence in young clusters. H$\beta$ emission can be seen, even when the young population only makes up 1-2\% of the total mass.  We take this as the quantitative limit of the present survey, at which a presently forming young population could be readily detectable.

We can make an additional test based on the example of T661 in NGC~3256 (\S~\ref{sec:t661}).  The V-band
flux ratio between the older ($50$~Myr) cluster and the H{\sc ii} region is $\sim3$\%, i.e. the older cluster is $\sim30$ times brighter.  It is difficult to turn this flux difference directly into a mass difference due to the unknown extinction towards the H{\sc ii} region.  However, unless if the extinction significantly higher towards the H{\sc ii} region, the implied mass difference is less than 1\%, in agreement with the estimates above (the mass-to-light ratio of 50~Myr cluster is $\sim7$ times that of a $\sim3$~Myr population, e.g., Bruzual \& Charlot 2003).

We have also investigated what effect stochastic sampling of the IMF has on the expected H$\beta$ luminosity.  To do this, we generated 100 realisations of clusters with masses of $10^2$, $10^3$ and $10^4$\msun, with a Chabrier (2003) IMF stochastically sampled, with stellar masses between 0.1 and 120\msun.  For each cluster, we calculated the total expected $H\beta$ luminosity based on the ionising photon rate of all of the member stars.  The ionisation rate as a function of teller mass was taken from the compilation of Davies et al.~(2011).  Since, to first order, we are only sensitive to secondary cluster formation with masses at or above 1\% of the older (initial) cluster mass, and that most of our clusters are $10^5$\msun\ or above, we focus on the $10^3 - 10^4$\msun\ simulations.

As an indicator, we use the relative difference between the mean total ionisation for all clusters (of a given mass) and the standard deviation, i.e. the relative scatter.   We find that the H$\beta$ luminosity for the $10^3$ and $10^4$\msun\ simulations have a relative scatter 1.5 and 1.15, respectively, in the 100 realisations.  Hence, for an original cluster of mass $10^5$\msun, in the worst case scenario this would move our detection limit for the secondary population to $\sim3$\% of the original cluster mass (2\% in the fully sampled case, multiplied by 1.5).  Hence, even in the low mass regime, we can place strong constraints on any ongoing star-formation within the cluster.

Finally, we can place a limit on the amount of ionised gas present within the clusters.  We assume Case B recombination, that the gas is fully ionised, and is composed soley of hydrogen.  From Storey \& Hummer~(1995) the emissivity of H$\beta$ (in the case that the electron temperature is less than 26000K) is Em(H$\beta$) = $10^{(-0.870 * log(Te) + 3.57)}$, which is related to the luminosity as L(H$\beta$) = $Em({\rm H}\beta) * n_{\rm e}^2 * volume$, where $n_{\rm e}$ is the election density.  For 100~\msun\ of ionised gas, at T$=10^4$~K, in a sphere of radius 1~pc, this leads to an expected luminosity of L(H$\beta$) $\sim 10^{37}$ ergs/s.  For a typical distance to a galaxy in our sample of 20~Mpc, this leads to an observed flux of $\sim 10^{-15}$ ergs/s/cm$^2$, which is about an order of magnitude brighter than the typical flux at H$\beta$ observed in clusters at this distance (e.g., Bastian et al.~2009).  Hence, we conclude that the clusters in our sample have $<100$~\msun\ of ionised hydrogen within them.

\section{Model Predictions and Constraints}
\label{sec:models}

\subsection{Extended SFHs in LMC/SMC clusters}

Observations of extended main-sequence turn-offs (eMSTOs) within intermediate age ($1-2$~Gyr) clusters in the LMC have been suggested to be due to extended star-formation events ($200-500$~Myr) within the clusters.  Conroy \& Spergel (2011) and Conroy (2012) have modelled such a system (see also the GC scenario outlined below) and suggest that a second population can form from a combination of processed material from AGB stars from the first generation along with pristine gas accreted from the surroundings.  In this model, gas builds up but does not form stars until the ionising flux rate of stars from the first population drops below a critical value.  The gas can then cool and form a second generation of stars.  This model predicts that there should be a 100-200~Myr delay between the first and second generation, with little or no star formation between the two bursts.

Goudfrooij et al. (2011a,b) and Keller et al. (2011) have suggested more empirically based conceptual models, where only massive clusters (i.e., those with high escape velocities and possibly those with large core radii) are able to retain gas within them in order to form successive generations of stars.  Goudfrooij et al. (2011b) suggest that only clusters with {\em initial} escape velocities above $10-15$~km/s can retain the stellar ejecta within the cluster and continue forming stars.  Here star formation is continuous during the first few hundred Myr of a cluster's life.  We note that the current escape velocities of the intermediate age LMC clusters are significantly lower than this critical value, and that cluster evolutionary models (that may invoke initial mass segregation) must be employed to extrapolate the initial conditions, with the majority of the change happening in the first few Myr of a clusters life.  Hence, in principle, we can compare the current estimated escape velocities of the clusters in our sample with the present day escape velocities of the LMC clusters.  However, we conservatively adopt the $15$~km/s escape velocity as a lower limit.  We also note that a number of the clusters in our sample are are significantly more massive than the intermediate age LMC/SMC clusters previously studied (by factors of $>10$).

The horizontal dashed lines in Fig.~\ref{fig:age_mass} show the mass limit of v$_{\rm escape} > 15$~km/s for cluster effective radii of 3, 5 and 10~pc, from bottom to top, respectively.  YMCs have typical effective radii of 2-5~pc (e.g., Larsen~2004; Scheepmaker et al.~2007), showing that the vast majority of the clusters in our sample are expected to have escape velocities well in excess of 15~km/s.  We note that this is their current escape velocities (and are much higher than the present day escape velocity in the intermediate age clusters) with their initial escape velocities potentially much higher (e.g., Goudfrooij et al. 2011b).  Hence, the prediction is that nearly all of the young clusters ($<200$~Myr) studied 
here should have ongoing star formation.

\begin{figure}
\includegraphics[width=8.5cm]{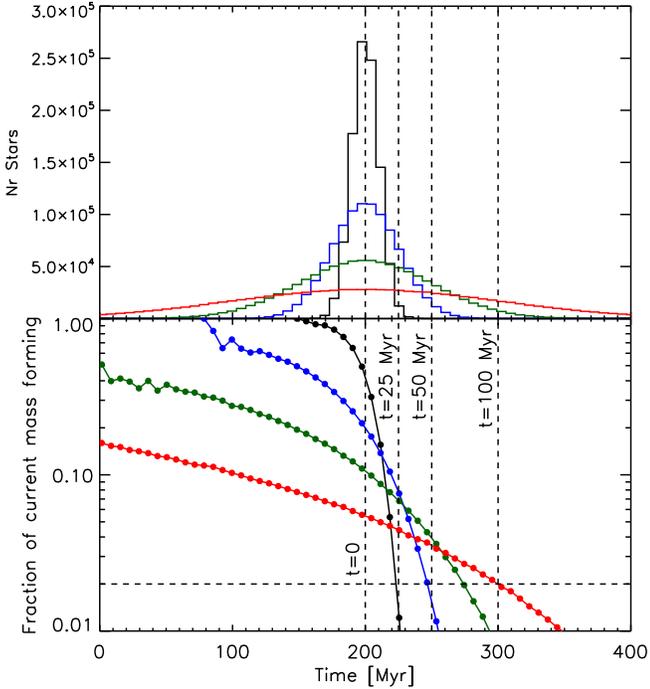}
\caption{{\bf Top panel:} The star-formation histories of four model clusters, adopting Gaussian distributions with dispersions of 10, 25, 50 and 100~Myr.  The peak of the distributions has been arbitrarily shifted to 200 Myr.  In this representation, star-formation begins on the left in the panel and time moves to the right.  {\bf Bottom panel:} The fraction of the current mass (the sum of all the stars that have formed until a given time) that is presently forming (in 7~Myr bins) in the cluster. t=0 represents the peak of the star-formation history and three times post-peak are indicated.  Note that if clusters are observed before the peak of the SFHs, they all have $>2$\% of their mass currently forming (and in some cases $>50$\%, depending on the duration of star-formation), hence would be expected to show (above the estimated detection limits) signs of ongoing star-formation.  }
\label{fig:extended_sfh}
\end{figure}

\begin{figure}
\includegraphics[width=8.5cm]{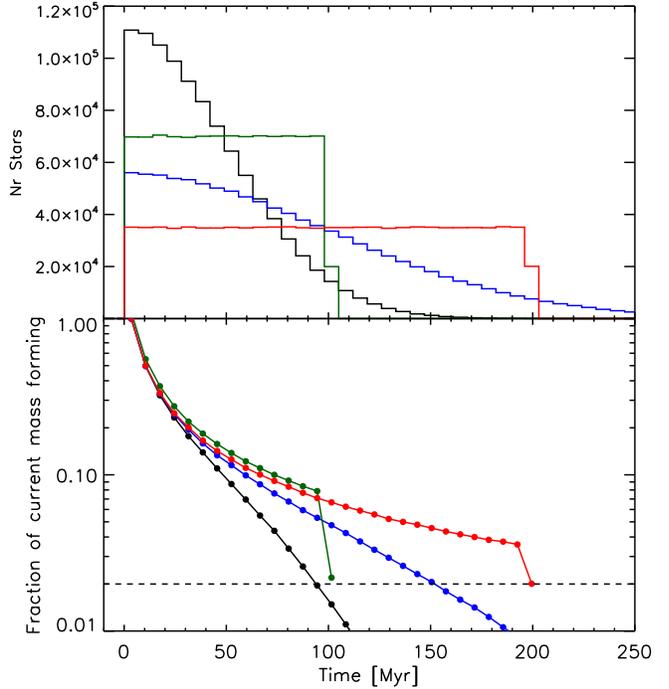}
\caption{Similar to Fig.~\ref{fig:extended_sfh} but for four different SFHs that all begin with a burst.   The red and green lines show constant SFHs of 200 and 100 Myr durations, respectively.  The blue and black lines are a bursts followed by a decay with a dispersion of 100 and 50~Myr, respectively.  In the bottom panel the green line has been shifted upwards by 10\% in order to separate it from the red line (any constant SFH would follow the same trend while star formation proceeds).}
\label{fig:extended_sfh2}
\end{figure}

In order to quantify the predictions, we generated clusters with $10^6$ stars, with SFHs sampled from a Gaussian distribution with four different dispersions, namely 10, 25, 50, and 100~Myr.  As a reference, the intermediate age clusters have been interpreted as having a Gaussian type SFH with a dispersion of 100-200~Myr (e.g., Goudfrooij et al. 2011b).  The SFHs are shown in the top panel of Fig.~\ref{fig:extended_sfh} where we have shifted the peak of the distributions arbitrarily to 200~Myr (in order to see the distributions before and after the peak).  We then calculate the fraction of the current cluster mass (i.e. the mass formed up to a given time) in 7~Myr bins (i.e. the approximate lifetime of H{\sc ii} regions) that is presently forming.  These distributions are shown in the bottom panel of Fig.~\ref{fig:extended_sfh}.  Note that this is not the fraction of the final mass of the cluster presently forming, but rather the fraction of the mass formed up to a given time.  Figure~\ref{fig:extended_sfh2} shows the same, but for four different SFHs.

For a randomly selected sample, we expect that half of the clusters would be seen before reaching the peak of their SFH.  Figure~\ref{fig:extended_sfh} shows that all such clusters would be presently forming a significant fraction of their mass.  Hence, we would expect that at least half of the clusters should show emission lines associated with ongoing star-formation.  Additionally, the actual fraction showing emission lines should be significantly more than 50\%, as ongoing star-formation is expected to be seen significantly beyond the peak of the SFH.

\subsection{Multiple generations of star formation in globular clusters}
\label{sec:mod_gc}

As discussed in \S~\ref{sec:intro}, multiple models have been put forward to explain the chemical anomalies observed in present day GCs.  Most of these models invoke a second burst of star formation, caused by the buildup of chemically processed material (and accreted pristine material) ejected from stars in the ``first generation".  Depending on the source of the enriched material, the timescale for the secondary burst can change.  If interacting binaries and/or rapidly rotating high mass stars are the origin, the second generation of stars is expected to form in the first $5-20$~Myr of a cluster's life.  Since our sample only contains clusters with ages greater than 10~Myr, we cannot constrain additional star forming events before this age. 


If AGB stars are the source of the pristine material, the second generation is expected to form between $30-200$~Myr, depending on the role of Lyman continuum photons in the cluster and when prompt Type Ia SNe begin to explode (e.g., D'Ercole et al.~2008, Conroy \& Spergel~2011).  These lower and upper limits are shown as vertical dash-dotted lines in Fig.~\ref{fig:age_mass}.  Conroy \& Spergel~(2011) estimate that all clusters with masses above $10^4$\msun\ in the LMC should undergo a second generation of star formation, after $100-200$~Myr. While the exact mass limit is environmentally dependent (due to the role of accretion and stripping onto/from the cluster), it is expected that many of the clusters in our sample should have (or have had) a secondary burst of star formation.  

D'Ercole et al.~(2008) do not specify an exact mass limit to when they expect multiple populations, only that the progenitors of present day GCs were $10-100$ times more massive than at present.  Since our sample has a number of clusters with masses in excess of $10^6$\msun\ (approximately 5-10 times more massive than the average GC) we may expect to observe the secondary burst of star formation.
 
We note that in the disc accretion model, no secondary star formation episode is expected (Bastian et al.~2013).  Here, the chemical enrichment of a fraction of the low-mass stars is due to the accretion of the ejecta of high mass stars and interacting binaries {\em of the same generation}.  Hence, a prediction of this model is that no clusters in the present sample should show signs of ongoing star-formation.

\subsection{Constraints from Observations}
\label{sec:results}

None of the clusters in the current survey display evidence of ongoing star formation within them.  This places strong constraints on both the possibility of extended star-formation periods within massive clusters (as proposed for intermediate age LMC and SMC clusters) and on the potential formation channels of the anomalies seen in globular clusters.

Firstly, we note that the lack of ongoing star formation is contrary to that expected for theories that invoke extended star-formation histories of clusters (Fig.~\ref{fig:extended_sfh}).  Goudfrooij et al.~(2011b) and Rubele et al.~(2013) suggest that star-formation continued for $>200$~Myr within many intermediate age LMC/SMC clusters.  In such a case, we would expect at least half of our sample (in particular, those with high escape velocities and/or high masses) to show strong emission lines, indicative of ongoing star formation.  However, all 130 young massive clusters in our sample lack any indication of ongoing star-formation within them.  The mean and median mass of the clusters in our sample are $2\times10^6$ and $2\times10^5$\msun, respectively.  In this mass range, even if a cluster was only forming only $10$\% of its current mass, the stellar IMF would be (nearly) fully sampled, so many high-mass stars would be expected to be present.

{\em We conclude that (semi)continuous star formation is not consistent with the observations,  hence that the observed eMSTOs and red clump morphologies of intermediate age clusters in the LMC/SMC are not caused by extended SFHs.}

Secondly, we look at the constraints on the proposed mechanisms for the anomalies observed in globular clusters.  The constraints imposed by the observations for the bursty/semi-continuous star-formation scenario are less stringent than for the continuous star-formation case, due to the finite sampling of age/mass space.  If the secondary star formation event is extremely short (e.g., $<1$~Myr) and only happens in massive (e.g., $>10^6$\msun) clusters, it is unlikely that such an event would be caught by our observations.  

\begin{figure}
\includegraphics[width=8.5cm]{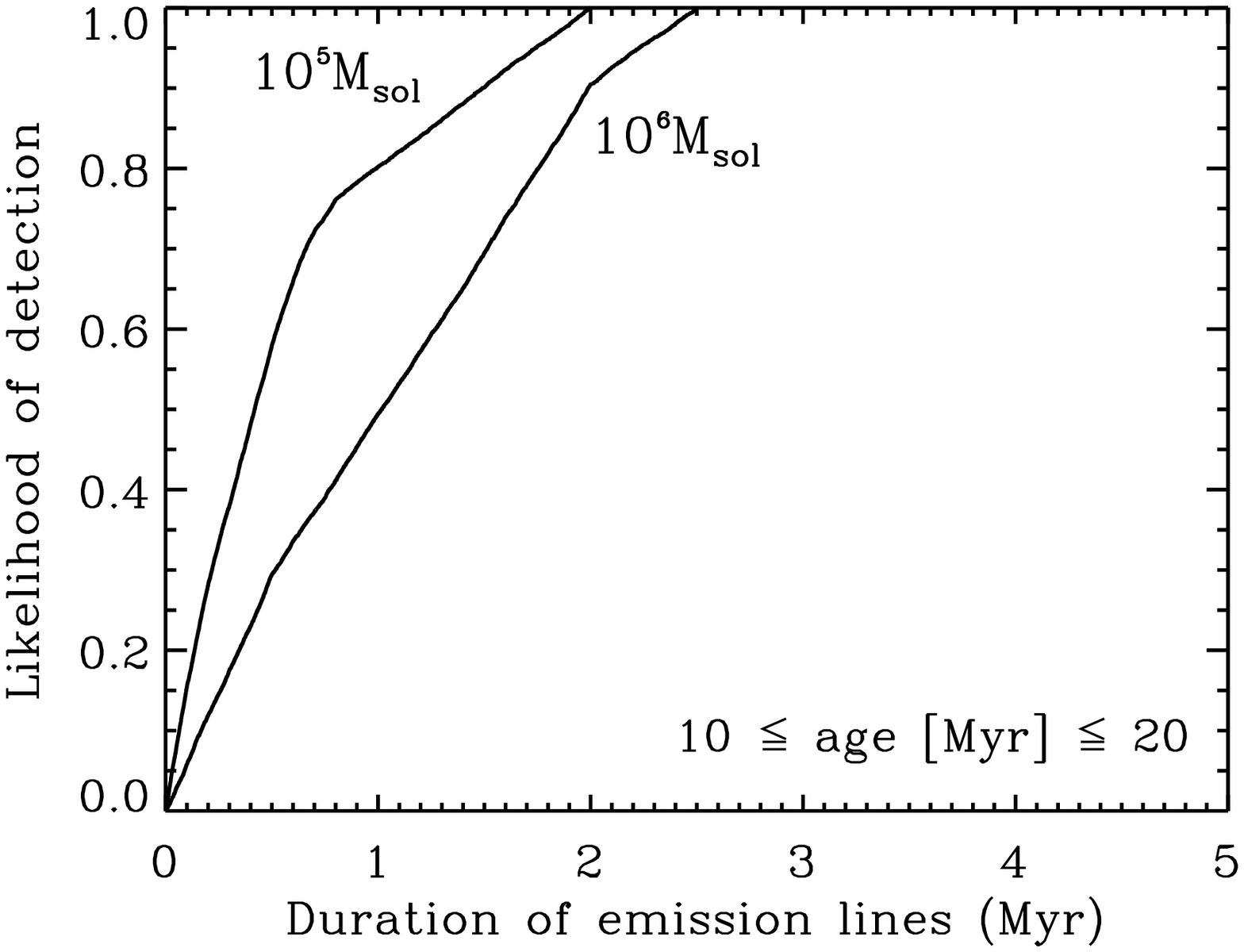}
\includegraphics[width=8.5cm]{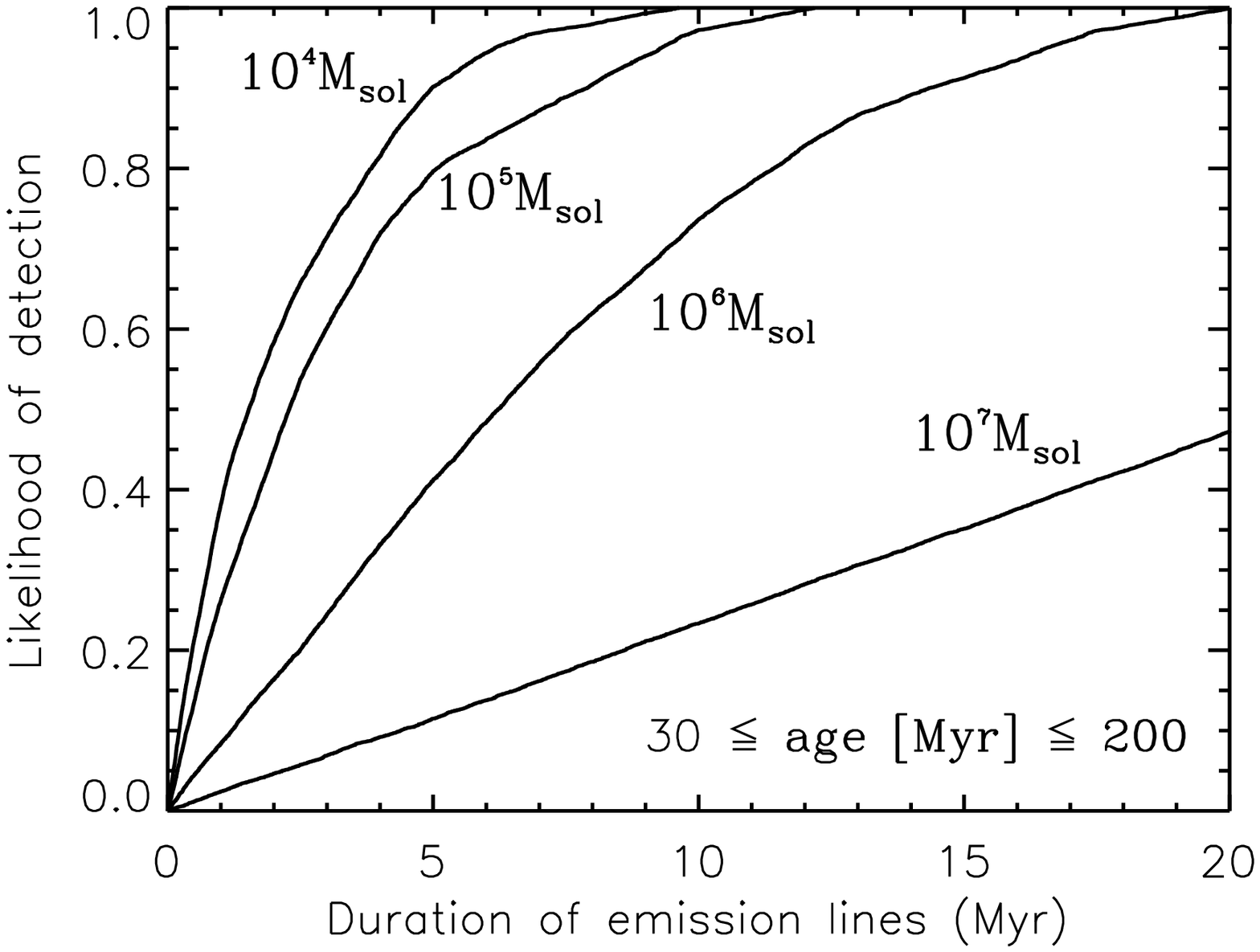}
\caption{{\bf Top panel: } The probability of detecting emission lines in a cluster forming a second generation of stars as a function of how long the emission lines last, given our sample of clusters with ages between $10-20$~Myr.  In normal H{\sc ii} regions, the duration of emission lines is $\sim7-10$~Myr. In order to assess the effect of cluster mass (i.e. if only clusters above a certain mass can form a second generation) we show two mass cuts, $10^5$\msun\ and $10^6$\msun, the upper and lower curves, respectively.  {\bf Bottom panel:} The same as the top but now focussed on clusters with ages between $30$ and $200$~Myr.  Four mass cuts are used, above $10^4$, $10^5$, $10^6$ and $10^7$\msun. }
\label{fig:mc}
\end{figure}

In order to quantify the precise constraints, we carried out two Monte-Carlo simulations.  The first focusses on clusters with ages between $10-20$~Myr, in order to test the models invoking the formation of a secondary population from the ejecta of massive stars.  The second focusses on the age range of 30-200~Myr, the predicted range for the second population in the AGB scenario.  In each Monte Carlo model, we stochastically generate 10000 clusters uniformly distributed in the given age range, and above a given mass cut, and look in our sample for the cluster with the nearest/closest age. In other words, we look at how well sampled the age-range is above a given mass cut.  Fig.~\ref{fig:mc} shows the results of the simulations.  The x-axis, ''the duration of emission lines", is the age difference between each simulated cluster and closest (in age) real cluster in our sample.  If the emission lines lasted this long (or more), the star-forming episode would have been detected in our sample.

We can place tight constraints on potential star formation between $10-20$~Myr, with near $100$\% detection rate if the emission lines are visible for at least 2~Myr and the mass limit above which clusters can form a second generation is at or below $10^5$\msun.  If the mass limit is $10^6$\msun, then we would have a $\sim75$\% chance of detecting such a secondary star formation event with emission lines lasting for 2 Myr.  If the emission lines last for 4~Myr, we would again have a near $100\%$ chance of detecting them.  However, we note that two of the most massive clusters in this age range have resolved photometry available, and neither show evidence for the presence of short-lived O-stars, providing further evidence against secondary star formation events over this timescale.

For the AGB scenario, which predicts that a second generation can form $30-200$~Myr after the formation of the first generation, we would have had a $\sim75$\% chance of detecting them if clusters with masses above $10^5$\msun\ could form them, if the emission would last for 5~Myr.  If a $10^6$\msun\ cluster is required to form a second generation (again taking 5~Myr duration as an example), then the probability would drop to $\sim40$\%.

In all cases, if the mass limit required to form a second generation was high (e.g., $>10^6$\msun) the probability of detecting the secondary episode becomes smaller.  This is due to the finite sampling of age-mass space, where there are fewer high mass clusters in our sample.  In all cases, we are sensitive to a secondary star forming episode that is $\sim1-2$\% of the mass of original cluster or more.

The observations presented here are consistent with the expectations of the "early disc accretion" model (Bastian et al.~2013), which predicts that clusters do not have secondary (or continuous) star-formation events.

The host galaxies of the clusters used in the current survey cover a wide range of galaxy morphologies.  They include quiescent and interacting spirals, dwarf starbursts, major and minor mergers, as well as post-starburst galaxies.   Observations of YMCs in galaxies have shown that there is a continuous distribution of cluster properties.  In particular, that there is a relation between the luminosity of the brightest cluster and the star-formation rate of the host galaxy (e.g., Larsen~2002), and that the number of clusters (above some luminosity cut) also scales with the SFR (e.g., Whitmore~2003).  This has been attributed to a size-of-sample effect, where a higher SFR results in more stars forming, which in turn results in more clusters, including the most massive clusters.  The Milky Way and LMC are currently forming (within the past 10 Myr) clusters with masses of about $10^5$~\msun\ (Westerlund~1 in the Galaxy and R136 in the LMC - Portegies Zwart et al.~2010) and galaxies like the Antennae with a factor of 10-20 times the Galactic SFR are forming clusters with masses up to a few times $10^6$\msun\ (Whitmore et al. 2010).  All the properties of the clusters seem to be the similar (characteristic size and general lifetimes), and the cluster mass functions within each galaxy are all broadly consistent with the form $Ndm \propto M^{-2}dm$ (with a potential exponential decline at high masses (see Portegies Zwart et al.~2010).  Due to these similarities and continuous distributions, there is little reason to suspect that the intermediate age clusters in the LMC/SMC are fundamentally different from those that we see forming today or any of those that formed in the past Gyr in nearby galaxies.

\section{Discussion and Conclusions}
\label{sec:discussion}

We have compiled a sample of 130 Galactic and extragalactic stellar clusters with ages between $10 \le {\rm age/Myr} \le 800$ and masses between $10^4 {\rm \le mass/\msun} \le 8 \times 10^7$ with published integrated spectroscopy or resolved photometry. 40 are presented here while 90 are taken from the literature.  In the present work, we search for evidence for any ongoing star-formation within the clusters, either through the detection of emission lines in the spectra (based on visual inspection) or for high mass stars through resolved photometry.  No clusters with ongoing star formation within them are found.  Upper limits can be placed on any ongoing star formation, and we find that if the forming population is $1-2$\% (or more) of the mass of the older population, it would be detected.  A handful of clusters with emission lines in the integrated spectra were found, however, upon closer inspection these were found to originate from nearby H{\sc ii} regions, hence were not related (at least directly) to the cluster under study (c.f., Peacock et al. 2013).

This lack of ongoing star formation is in stark contrast to what would be expected if clusters had extended star formation episodes of hundreds of Myr, as has been suggested for intermediate age clusters in the LMC/SMC (e.g., Mackey \& Broby-Neilsen~2007; Mackey et al.~2008; Milone et al.~2009, Goudfrooij et al.~2009, 2011a,b; Rubele et al. 2013).  The majority of clusters in the sample have escape velocities that are expected to be well in excess of the limit of $10-15$~km/s suggested by Goudfrooij et al.~(2011b) as necessary for extended star-forming periods.   The lack of ongoing star formation within young clusters, and the lack of extended SFHs in resolved young massive clusters in the LMC (Bastian \& Silva-Villa~2013)  are seemingly at variance with the age-spreads inferred from the extended main sequence turn-offs and red-clump morphologies observed in intermediate-age clusters in the LMC/SMC. This may indicate that alternative explanations are needed for these phenomena.

Similarly, the AGB scenario for GCs, where processed material from AGB stars is retained within the cluster and (when mixed with pristine material from the surroundings) can form a second generation of stars, also predicts that a secondary burst of star formation should be present within young massive clusters.  While we do not find evidence for this, the observations presented here cannot definitively rule out such a scenario.  If the second generation forms extremely rapidly (in less than a few Myr) and no gas is left in or around the cluster, it is possible that the star forming event may have been missed, given our finite sampling of age/mass space.  However, in a future work, Cabrera-Ziri et al. (in prep), we will analyse the integrated spectra of a sub-sample of these clusters in detail (focussing on the highest mass clusters), in order to place direct constraints on their SFHs.

One potential caveat to the current analysis is that if the stellar initial mass function of the second generation was devoid of high mass stars capable of producing ionising photons, then no line emission would be seen in the integrated spectra.  However, there is little evidence for a strongly varying IMF, especially in massive clusters (e.g., Bastian, Covey \& Meyer~2010).  Many resolved young clusters are seen with high mass stars (e.g., Clark 
\& Negueruela~2002; Crowther et al~ 2010) and young ($<7$~Myr) high mass extragalactic clusters ($10^5 - 10^6$\msun) show strong Wolf-Rayet emission features, showing that such stars can form/exist within dense clusters (e.g., Trancho et al.~2007b; Sidoli et al.~2006 - at least in the first generation).  So, there does not seem to be any reason, {\em a priori}, that stellar populations within clusters should be devoid of high-mass stars.

In order to fully constrain whether existing stellar clusters can host secondary star formation events (with the exception of nuclear clusters - which are known to host multiple episodes of star formation - e.g., Rossa et al. 2006, Seth et al. 2008) it is necessary to derive their full SFH from integrated spectroscopy.  While challenging, it is within reach using high S/N medium resolution spectra, due to the rapidly changing SED and spectral features of young stellar populations.  We will present the tools and the derived constraints in an upcoming work, Cabrera-Ziri et al.~(in prep.).


\begin{table*} 
\caption{Cluster ID, galaxy, age, mass, and associated comments for the clusters used in the present survey.  See \S~\ref{sec:catalogue} for a description of the data and estimated errors.  The references are labelled as follows:  1 - Bastian et al.~(2008); 2 - Konstantopoulos et al.~(2009); 3 - Gallagher \& Smith~(1999); 4 - Smith \& Gallagher~2001); 5 - McCrady et al.~(2007); 6 - McCrady, Gilbert \& Graham~(2003); 7 - Bastian et al.~(2009); 8 - Trancho et al.~(2012); 9 - Trancho et al.~(2007a); 10 - Trancho et al.~(2007b); 11 - Schweizer \& Seitzer~(1998); 12 - Maraston et al.~(2004); 13 - Bastian et al.~(2006); 14 - Schweizer \& Seitzer~(2007); 15 - Larsen, Brodie \& Hunter~(2004); 16 - Larsen \& Ritchler~(2004); 17 - Chien et al.~(2007); 18 - Larsen \& Richtler~(1999); 19 - Bastian et al.~(2012); 20 - Larsen \& Brodie~(2002); 21 - Brodie et al. (1998); 22 - Schweizer, Seitzer \& Brodie~(2004); 23 - Caldwell et al. (2009); 24 - de Grijs et al.~(2004); 25 - Moll et al.~(2007); 26 - Bastian et al.~(2006); 27 - Whitmore et al.~(2010); 28 - Bastian \& Silva-Villa~(2013); 29 - Larsen et al.~(2011); 30 - Larsen et al.~(2008); 31 - V{\'a}zquez et al.~(2004); 32 - Figer et al.~(2006); 33 - Davies et al.~(2007); 34 - Sirianni et al.~(2002); 35 - McLaughlin \& van der Marel~(2005); 36 - Larsen~(2004); 37 - Larsen~(2009); 38 - this work
}
  \begin{tabular}
    {lclcccc}
    \hline
 ID  & Galaxy &  age (Myr)  & mass ($10^5 \msun$)& reference & comments\\
    \hline 
3cl-a& M51 &	   20&	   	    1.0	      &  1                        &          \\
34&   M82 &	   90&		    3.0	      &  2			       &	  \\
34.1& " &  	   160&		    14     &  2			       &	  \\
43.2& " &  	   130&		    1.4     &  2			       &	  \\
F (51)&    "   &	   70&		    5.5     &  2,3,4,5		       & see \S~\ref{sec:m82f}	  \\
L&    "   &	   65&		    40     & 2,3,4,5		      	       & see \S~\ref{sec:m82f}	  \\
78.2& "  &	   20&		    3.0	      & 2			       &	  \\
H (91)&" &	   200&		    20	      & 2			       &	  \\
97&   "  &	   160&		    3.0	      & 2			       &	  \\
103.2&"  &	   80&		    2.0	      & 2			       &	  \\
108&  "  &	   160&		    2.4     & 2			       &	  \\
29&   "  &  	   80&		    1.4     & 2			       &	  \\
34&   "   &	   90&		    2.9     & 2			       &	  \\
58&   "  &	   30&		    0.46     & 2			       &	  \\
78.1& "  &  	   30&		    0.16     & 2			       &	  \\
103.2&"  &	   80&		    2.0	      & 2			       &	  \\
131&  "   &	   80&		    3.8     & 2			       &	  \\
140&  "   &	   60&		    0.75     & 2			       &	  \\
147&  "   &	   70&		    0.46     & 2			       &	  \\
MGG-9&" &  	   10&		    15     & 6			       &	  \\
MGG-11&" &	   10&		    3.5     & 6			       &	  \\
T111&NGC~4038/39&  85&		    2.0	      & 7			       & see also Peacock et al. (2013) \\
T313&	"   &63&		    1.0	      & 7			       &	  \\
T296&	"   &80&		    4.0	      & 7			       &	  \\
T299&	"   &23&		    2.5	      & 7			       &	  \\
T130&	"   &250&		    1.0	      & 7			       &	  \\
T395&	"   &630&		    2.0	      & 7			       &	  \\
T297&	"   &300&		    1.6	      & 7			       &	  \\
T118&  	Stephan's Quintet   &126&   20	      & 8			       &	  \\
T120&	"   &13&		    3.0	      & 8			       &	  \\
T121&	"   &40&		    1.3     & 8			       &	  \\
T113&	"   &50&		    0.9     & 8			       &	  \\
T119&	"   &13&		    6.7     & 8			       &	  \\
T122&	"   &50&		    0.2     & 8			       &	  \\
T123&	"   &10&		     2.0    & 8			       &	  \\
T1127&   NGC~3256   &80&	    1.5     & 9			       &	  \\
T1149&   "   &80&		     1.6     & 9			       &	  \\
T1165& 	"   &200&		     2.8    & 9			       &	  \\
T88&   	"   &30&		     1.5    & 10			       &	  \\
T99&	"   &30&		     22    & 10			       &	  \\
T112&	"   &100&		     110     & 10			       &	  \\
T661&	"   &50&		     13    & 10			       & see \S~\ref{sec:t661}	  \\
T1078&	"   &160&		     14    & 10			       &	  \\
W3& NGC~7252	   &400&	     800      & 11,12		       &	  \\
W30& 	"   &400&		     180    & 11,13		       &	  \\
W6&	"   &400&		     150    & 11			       &	  \\
W26&	"   &400&		     76    & 11			       &	  \\
Cluster 1& NGC 34&150&		     150     & 14			       &	  \\
Cluster 2& 	"   &160&	     100     & 14  & used younger age    	  \\
10 & NGC 4214 & 200&		     26    & 	   15		       &	  \\
13 & NGC 4449 & 200&		     15    &	   15		       &	  \\
27 & " & 800&		             21     &	   15		       &	  \\
47 & " & 280&		     	     46    &	   15		       &	  \\
11 &NGC 6946  & 11&		     18    &	   15		       &	  \\
    \hline 
  \end{tabular}
\label{tab:clusters}
\end{table*}

\setcounter{table}{0}

\begin{table*} 
\caption{continued.
}
  \begin{tabular}
    {lclcccc}
    \hline
 ID  & Galaxy &  age (Myr)  & mass ($10^5 \msun$)& reference & comments\\
    \hline 
502 & M83    &100&		     5.0      &	   16		       &	  \\
805 & M83	   &13&		     2.0      &	   16		       &	  \\
Cluster 1 & NGC~4676 &170&	    10      &	   17		       &	  \\
L679& NGC~2997 &15.8&  		     4.8    &   	   18,19	       &	  \\
L693&	"   &13 & 		     3.4    &	   18,19	       &	  \\
L715&	"   &16  &		     2.9    &	   18,19	       &	  \\
L836&	"   &42  &		     3.8    &	   18,19	       &	  \\
L645&	"   &79  &		     4.3    &	   18,19	       &	  \\
L405&	"   &13  &		     1.1    &	   18,19	       &	  \\
L696&	"   &32  &		     0.97    &	   18,19	       &	  \\
L648&	"   &63  &		     1.5    &	   18,19	       &	  \\
L690&	"   &25  & 	     	     0.41    &	   18,19	       &	  \\
3 & NGC 1023A    & 300	     &	      0.8  &	20	       	\\
4 & NGC 1023A    & 300	     &	      0.6  &	20		\\
H1 & NGC~1275	 & 400      &	      200  &    21 	      & the emission lines in the spectrum offset by $>100$~km/s\\
S1 & NGC~3921	 & 450      &         23   & 22 & \\
S2 & NGC~3921    & 280      & 	      15   & 22 & \\
VDB0 & M31	 & 40	    &	      1.3  & 23 & many more clusters are available, however these are the most\\
B315-G038 & M31  & 130	    &	      1.0  & 23 &  massive and have published spectroscopy\\
B367-G292 & M31  & 200	    &	      0.2  & 23 & "\\
B049-G112 & M31	 &  440	    &	      0.4  & 23 &" \\
B314-G037 & M31	 &  550	    &	      0.13 & 23 & "\\
Cluster 6 & NGC~1140 & 20   &	      91   & 24,25 & age from integrated photometry\\
w37254 & NGC~4038/39 & 30 & 16 & 26,27 &  ", emission lines offset by $\sim135$~km/s, complex 3 \\
\hline
NGC 1856 & LMC &280   &		     0.76     &	   28		       & based on resolved photometry	  \\
NGC 1866 & " &180&		     0.81     &	   28		       &      "	  \\
B & NGC~1569&16&		     12       &    29,30	   	       & ", spectroscopy also available	  \\
1 & NGC~1705&13&		     9.2      &	   29,31		       & ", spectroscopy also available	  \\
F3-1& NGC~1313&55&		     2.8      &	   29		       & based on resolved photometry	  \\
F1-1& NGC~5236&27&		     2.1      &	   29		       & "	  \\
F1-3& NGC~5236&28&		     8.1      &	   29		       & "	  \\
F1-1& NGC~7793&50&		     0.92     &	   29		       & "	  \\
RSGC1 & Milky Way &12&		     0.5      &	   32		       & "	  \\
RSGC2 & Milky Way &17&		     0.64     &    33		       & "	  \\
NGC 330 & SMC     & 30  &            0.5      &    34, 35              & "        \\                         
\hline
622       &          NGC~2835     &         81   &      0.66		         & 36,37,38	  &   ages derived from integrated photometry     \\
376       &          NGC~2997     &        181   &      1.92		         &   "	  &     "   \\
1275      &          NGC~3621     &         14   &      0.76		         &   "	  &     "   \\
1032      &            "          &        181   &      1.28		         &   "	  &     "   \\
517       &            "          &        102   &      1.18		         &   "	  &     "   \\
793       &            "          &         14   &      1.48		         &   "	  &     "   \\
630       &            "          &        509   &      8.24		         &   "	  &     "   \\
512       &            "          &        128   &      0.97		         &   "	  &     "   \\
1704      &          NGC~5236     &         40   &      0.68		         &   "	  &     "   \\
1630      &             "         &        361   &      1.80		         &   "	  &     "   \\
1540      &             "         &         72   &      0.53		         &   "	  &     "   \\
1471      &             "         &         57   &      0.87		      &   "	  &     "   \\
1395      &             "         &        161   &      0.77		      &   "	  &     "   \\
1321      &             "         &        143   &      0.72		      &   "	  &     "   \\
1253      &             "         &        181   &      0.51		      &   "	  &     "   \\
1182      &             "         &         13   &      0.98		      &   "	  &     "   \\
854       &             "         &         34   &      1.16		      &   "	  &     "   \\
429       &             "         &        203   &      0.41		      &   "	  &     "   \\
367       &             "         &         72   &      1.05		      &   "	  &     "   \\
288       &             "         &        181   &      1.89		      &   "	  &     "   \\
245       &             "         &        102   &      1.36		      &   "	  &     "   \\
1403      &             "         &        361   &      1.09		      &   "	  &     "   \\
\hline
  \end{tabular}
\label{tab:clusters}
\end{table*}

\setcounter{table}{0}

\begin{table*} 
\caption{continued.
}
  \begin{tabular}
    {lclcccc}
    \hline
 ID  & Galaxy &  age (Myr)  & mass ($10^5 \msun$)& reference & comments\\
    \hline 
1273      &             "         &        102   &      0.49		      &   "	  &     "   \\
944       &             "         &        128   &      1.51		      &   "	  &     "   \\
732       &             "         &         47   &      0.72		      &   "	  &     "   \\
548       &             "         &        181   &      1.26		      &   "	  &     "   \\
515       &             "         &         31   &      0.24		      &   "	  &     "   \\
240       &             "         &        405   &      1.27		      &   "	  &     "   \\
218       &             "         &        361   &      0.21		      &   "	  &     "   \\
126       &             "         &        321   &      0.97		      &   "	  &     "   \\
1227      &             "         &        181   &      0.47		      &   "	  &     "   \\
1187      &             "         &        286   &      0.57		      &   "	  &     "   \\
821       &             "         &        114   &      0.49		      &   "	  &     "   \\
676       &             "         &        228   &      0.47		      &   "	  &     "   \\
599       &             "         &        286   &      1.06		      &   "	  &     "   \\
578       &             "         &         52   &      0.48		      &   "	  &     "   \\
360       &             "         &         45   &      0.47		      &   "	  &     "   \\
310       &             "         &        161   &      0.82		      &   "	  &     "   \\
236       &             "         &        405   &      0.33		      &   "	  &     "   \\
693	  &		" 	  &	   36	 &      0.60        &   "	  & H{\sc ii} region nearby\\

    \hline 
  \end{tabular}
\label{tab:clusters}
\end{table*}

\pagebreak

\section*{Acknowledgments}
We thank Barbara Ercolano, Steve Zepf, and Iraklis Konstantopoulos for insightful discussions as well as Fabio Bresolin for helpful discussions and for providing H{\sc ii} region spectroscopy for comparison.  NB  is partially funded by a University Research Fellowship from the Royal Society.  IC-Z acknowledges the support from a postgraduate grant from CIDA. BD is partially funded by a Royal Astronomical Fellowship. 

\end{document}